\documentclass[sigconf]{acmart}

\usepackage{booktabs} % For formal tables
\usepackage{hyperref}
\usepackage{graphicx}
\usepackage{subfig}
\usepackage{amsthm}
\usepackage{verbatim}
\usepackage[bottom]{footmisc}
\graphicspath{ {figures/} }

\setcopyright{rightsretained}
\acmConference[SIGIR 2018]{ACM SIGIR conference}{July 8-12}{En Harbor, Michigan USA} 
\begin{document}
\title{Preliminary Bias Results in Search Engines}
\author{Gizem Gezici}
\affiliation{%
  \institution{Huawei Turkey R\&D Center}
  \city{Istanbul}
  \country{Turkey}}
\email{gizem.gezici@huawei.com}

%\titlenote{Produces the permission block, and
  %copyright information}
%\subtitle{Sentiment-wise Comparison of Bing \& Google}
%\subtitlenote{The full version of the author's guide is available as
  %\texttt{acmart.pdf} document}

\begin{abstract}
This report aims to report my thesis progress so far. My work attempts to show the differences in the perspectives of two search engines, Bing and Google on several selected controversial topics.
%In the first part, we try to make a link between the instances with high disagreement and the accuracy of the task on a crowd labelling environment. We claim that the instances to be labelled by annotators can be chosen and the budget can be allocated to the instances which can provide more information to the system. In this way, we do not waste budget and exploit the crowd labellers in a smarter way to increase the performance of the given task. 
In this work, we try to make a distinction on the viewpoints of Bing \& Google by using sentiment as well as the ranking of the document returned from these two search engines on the same queries, these queries are related mainly to controversial topics. You can find the methods we used with experimental results below.

% paragraph may be revisited later!

\end{abstract}

\begin{comment}

\begin{abstract}
This report aims to measure bias in Web search engines. For this, we analyse the search results of two commonly used search engines, Bing and Google in terms of sentiment on document, sentence and aspect-level. In order to find clues of bias (if any), we propose two metrics that blend sentiment and document ranking information together. Then, by using these metrics we compare two search engines, Bing and Google on 15 controversial topics. From the results, two general conclusions can be drawn: i) None of these search engines shows an extreme behaviour i.e. always positive or negative, ii) Only on specific controversial topics, these two search engines behave significantly different, i.e. obviously favoring opposite poles in plots. These conclusions do not advocate the claim of bias in search engines; by looking at these results one cannot defend that it exists to the extent that it affects the users of these search engines.
% paragraph may be revisited later!

\end{abstract}

\end{comment}


%\ccsdesc[500]{Computer systems organization~Embedded systems}

%\keywords{ACM proceedings, \LaTeX, text tagging}

\maketitle

\section{Introduction}
%My progress so far is mainly composed of two parts. In the first part, I have studied in collaboration with my thesis advisor's another PHD student, Stefan Rabiger on crowd sourcing. The aim of this part is to propose a crowd labelling framework for an effective budget usage, i.e. prioritizing the labelling of specific instances that are to be better labelled by crowd workers. I will give more details about the proposed crowd labelling framework in the next section. Note that the first part of my thesis, which I am the co-author, is a complete work and it has just been sent for a conference, whereas the second part has not been finalized, yet and I will only report the results so far. 

The core component of my thesis work and our main goal is to show the existence of bias in search engines (Bing \& Google) which is a very strong claim, therefore we started to differentiate the attitudes of these two search engines on some controversial topics. You can find the details of our method with the experimental results we obtained in the following sections after the overview of our \textit{unpublished work}, effective crowd labelling procedure.

\section{Sentiment-wise Comparison of Two Search Engines}
Two popular search engines were mainly compared from sentiment perspective by exploiting three different levels of sentiment metrics such as \textit{document-level}, \textit{sentence-level}, and \textit{aspect-level}.. Initially, we computed sentiment scores by using TextBlob as well as the polarity values of the words in a widely used domain-independent lexicon, SentiWordNet \cite{baccianella2010sentiwordnet}. However, obtained values from SentiWordNet were not reasonable and did not reflect the real sentiment of the given document. Therefore, we decided to use only TextBlob for sentiment polarity detection in our experiments.
\subsection{Document-level Sentiment Analysis} For document-level analysis, we obtain a sentiment value for each document from TextBlob without any pre-processing step on documents, TextBlob handles the punctuation in itself. This part of the analysis is more coarse-grained relatively to the subsequent parts.

\subsection{Sentence-level Sentiment Analysis} Differently from 
document-level, in sentence-level sentiment analysis we view each document as a pile of sentences. This leads to a more fine-grained analysis; thus in this part, we split the given document into sentences and obtain polarity value for each of its sentences from TextBlob. In order to obtain a sentiment score for the given document, we compute an  average sentiment by summing the polarity values of all the sentences and dividing it by the number of sentences in the document. That is to say, from TextBlob's point of view, each sentence is a document and processed accordingly. For clarification, again we have one polarity value for each document to compare the retrieved document sets of two search engines, although we compute these scores in distinct ways.

\subsection{Aspect-level Sentiment Analysis} In addition to document and sentence-level, we also analyzed the documents on aspect-level with the hope that it may convey different type of information that is useful for the comparison of the search engines. In analyzing the documents from the aspect-level perspective, we implemented a Java script on Eclipse environment in order to find dependencies between words. In this script, we read documents, split the documents into sentences and with the help of Stanford NLP Parser \cite{stanfordnlpparser} we obtain relations between words. In this way, grammatically we can have deeper understanding of documents and this additional information is exploited while analyzing the documents in terms of sentiment. As a side note, detailed information about the grammatical relations can be found in the Stanford NLP Parser manual\footnote{\url{https://nlp.stanford.edu/software/dependencies_manual.pdf}}.

In extracting dependencies between words in sentences, we only pay attention to the relations of the controversial topic keywords. We select these keywords for each topic as follows, if the given topic is composed of one word, we use that word as the keyword for the corresponding topic. If the topic contains more than one word, then we tokenize the topic title and all of the tokens are used as the keywords for that topic. Note that in our current dataset, topic titles are composed of at most two words, thus we have to deal with two-keywords case in our dependency analysis differently from the simple case of one-keyword topics. You can find sample relations extracted from the one-keyword controversial queries \textit{abortion} and \textit{brexit} in Table \ref{table:depAbortBrexit}. For two-keywords queries, we obtain dependencies for both of the keywords in the given topic and use all of these dependencies for the analysis. You can find sample relations extracted from the two-keywords controversial query \textit{gay marriage} in Table \ref{table:depGayMarriage}.

While extracting relations between words, initially we find keyword/s for the given topic, then we merely extract the relations of those keyword/s as mentioned above. Apart from this, we obtain the relations between keywords and only sentiment-baring words i.e., nouns, adjectives, adverbs, verbs. This is stemmed from the fact that we use dependency information for sentiment analysis and non sentiment-baring words will not be useful in our analyses. After getting relations from the parser, we will compute the sentiment score of the words that have grammatical relations with the keywords of the given topic. Sentiment polarities are computed by using TextBlob for all the related words in a document, then we compute an average sentiment score of each document for comparison.

%Each relation has three members as two words and the relation between these words. Depending on the relation, one of these words will affect, the other one will be affected from that word. This subject-object relation information is obtained from the Stanford NLP Parser that one word will be labelled as \textit{governor}, and the second one as \textit{dependent} with the specific relation name between these words. In our current analyses, we did not use this governor-dependent information and report the dependencies as the second word is the keyword of the given topic. 

\begin{table}
\begin{center}
\begin{tabular}{ |c|c|c| } 
 \hline
 \textbf{Word-Relation-Keyword}  \\ 
 \hline
 demand - nmod:on - abortion \\
  \hline
illegal - nsubj - abortion \\
 \hline
laws - compound - abortion\\
 \hline
ban - dobj - abortion\\
 \hline
restrict - dobj - abortion\\
 \hline
  \hline
  negotiations - compound - brexit\\
 \hline
bill - compound - brexit\\
 \hline
risk - nmod:of - brexit\\
 \hline
chaotic - amod - brexit\\
 \hline
 doubts - compound - brexit \\
  \hline
\end{tabular}
\end{center}
\caption{Sample Dependencies of \textit{Abortion} \& \textit{Brexit}}
\label{table:depAbortBrexit}
\end{table}

\begin{table}
\begin{center}
\begin{tabular}{ |c|c|c| } 
 \hline
 \textbf{Word-Relation-Keyword}  \\ 
 \hline
 rights - amod - gay \\
  \hline
community - amod - gay \\
 \hline
activist - amod - gay\\
 \hline
fans - amod - gay\\
 \hline
nice - amod - gay\\
 \hline
  \hline
  legalizing - dobj - marriage\\
 \hline
support - nmod:for - marriage\\
 \hline
recognize - dobj - marriage\\
 \hline
outlaw - dobj - marriage\\
 \hline
opposed - dobj - marriage \\
  \hline
\end{tabular}
\end{center}
\caption{Sample Dependencies of \textit{Gay Marriage}}
\label{table:depGayMarriage}
\end{table}

\section{Experimental Results}
\subsection{Dataset}
In order to observe remarkable differences between the retrieval 
results of two popular search engines, we selected 15 controversial topics from the website of \url{https://www.procon.org/}. We collected data using news API of the corresponding search engines: Bing and Google. With the aim of retrieving more documents related to each topic, we extended the query set by using Google Trends. Then, we implemented scripts on python to automate the crawling process with the API keys of the search engines. 

For each query sent to the search engine, we crawled the first 10 documents returned from that search engine. The dataset is composed of 2560 documents in total for one search engine. Topic distribution of the crawled dataset is displayed in Table \ref{table:topdist}. In the table, one can see that there are 200 documents crawled for the controversial query of \textit{abortion}, this means that we have 20 extended queries for this specific topic since we use only the first 10 documents retrieved from the corresponding search engine. We note that the data collection process was fulfilled in a controlled environment such that the same queries were sent to the search engines almost at the same moment since the documents to be retrieved may vary by time.

\begin{table}
\begin{center}
\begin{tabular}{ |c|c|c| } 
 \hline
 \textbf{Topic} & \textbf{\# of documents} \\ 
 \hline
 Abortion & 200 \\ 
 \hline
 Animal Testing & 80  \\ 
 \hline
 Assisted Suicide & 80 \\ 
 \hline
 Brexit & 180 \\ 
 \hline
 Climate Change & 350 \\ 
 \hline
 Gay Marriage & 220 \\ 
 \hline
 Gun Control & 260 \\ 
 \hline
 Medical Marijuana & 130 \\ 
 \hline
 Minimum Wage & 220 \\ 
 \hline
 Obamacare & 120 \\ 
 \hline
 Prostitution & 70 \\ 
 \hline
 Syrian Refugees & 130 \\ 
 \hline
 Transgender Military & 80 \\ 
 \hline
 Travel Ban & 140 \\ 
 \hline
 Trump & 300 \\ 
 \hline
\end{tabular}
\end{center}
\caption{Controversial Topic Distributions}
\label{table:topdist}
\end{table}

\subsection{Evaluation Framework}
In evaluating two search engines, our aim is to compare these two search engines on conservative/liberal perspective towards a given controversial
topic. For comparison, we utilize TextBlob sentiment scores at three different levels , \textit{document}, \textit{sentence}, and \textit{aspect-level}. 

In order to make a consistent comparison among all controversial topics, one needs to think about the connection between the semantic orientation and the perspective of the given documents on the scope of their controversial topics. For instance, if a document has a positive sentiment score and its controversial topic is \textit{abortion}, then the document's perspective tends to be more liberal, whereas if a document shows a negative attitude towards \textit{brexit}, similarly its perspective would be more close to liberal. This is because of the fact that the underlying perspective of a document on conservative/liberal axis in our evaluation scheme does not only depend on its semantic orientation but also on the corresponding controversial topic. For this reason, after obtaining polarity values at three different levels via TextBlob, we transformed these values for some of the controversial topics in order to make a consistent comparison for all controversial topics. In order to fulfill the transformation task, the polarity values were simply multiplied by -1 for \textit{animal testing}, \textit{brexit}, \textit{minimum wage}, \textit{travel ban}, and \textit{trump}. 

In our evaluation scheme, we used two metrics and computed these at three different levels, \textit{document}, \textit{sentence}, and \textit{aspect-level}. As the first metric, sentiment scores obtained from TextBlob were used after the transformation process as mentioned above, if necessary. Secondly, we proposed a new metric that combines rank and sentiment information and takes into account both of these for comparison. Also for the second metric, we used transformed sentiment scores to provide a consistent perspective analysis, instead of the actual polarity values obtained from TextBlob.

\subsubsection{Sentiment Polarity Metric}
\begin{figure}
\centering
  \includegraphics[width=80mm]{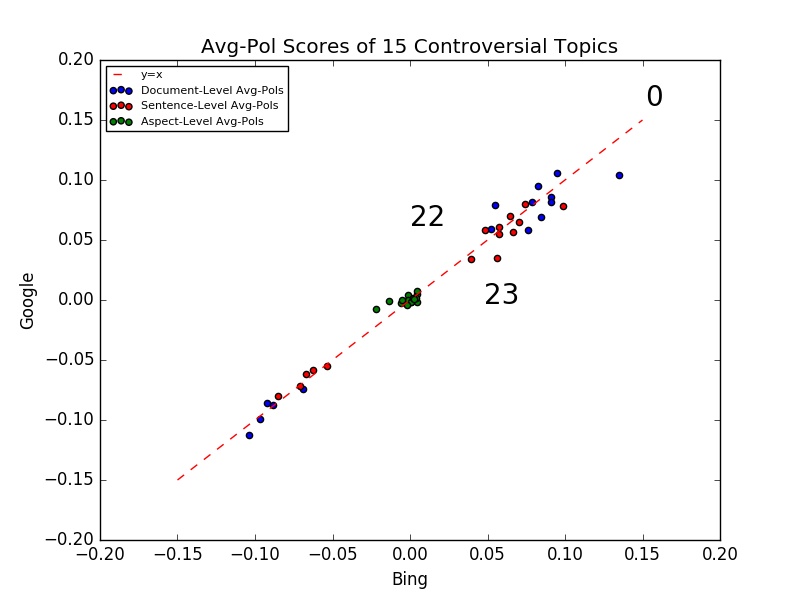}
  \caption{Mean Avg-Pol Scores of 15 Controversial Topics}
  \label{fig:avgPolScore}
\end{figure}

Our first metric compares the two search engines by purely using sentiment scores returned from TextBlob and the range for polarity values is [-1, 1]. We obtained sentiment scores for documents, sentences, and aspects, then computed an average polarity value as mentioned in the previous section for each document by exploiting \textit{document}, \textit{sentence}, and \textit{aspect-level} sentiment values separately. That means that we only care about document sentiment scores for comparison purposes, yet by computing these document scores in three different levels of sentiment analysis then transforming some of these polarity values if needed. You can find the table of mean of mean average sentiment scores in Table \ref{table:meanAvgPol} and comparison plot of two search engines in terms of mean avg polarity scores in Figure \ref{fig:avgPolScore}.

\begin{table}
\begin{center}
\begin{tabular}{ |c|c|c|c| } 
 \hline
 \textbf{Sentiment-level} & \textbf{Bing}  & \textbf{Google}\\ 
 \hline
 Document-level & 0.0260 &	0.0241 \\ 
 \hline
 Sentence-level & 0.0195 &	0.0178 \\ 
 \hline
 Aspect-level & -0.0018 &	0.0001 \\ 
 \hline
\end{tabular}
\end{center}
\caption{Mean of Mean Average Sentiment Scores over All Controversial Topics}
\label{table:meanAvgPol}
\end{table}

\begin{table}
\begin{center}
\begin{tabular}{ |c|c|c|c| } 
 \hline
 \textbf{Sentiment-level} & \textbf{Bing}  & \textbf{Google}\\ 
 \hline
 Document-level & 0.7845 &	0.7916 \\ 
 \hline
 Sentence-level & 0.7821 &	0.7967 \\ 
 \hline
 Aspect-level & 0.6099 &	0.5234 \\ 
 \hline
\end{tabular}
\end{center}
\caption{Mean Average \textit{NDCG-Senti} Scores over All Controversial Topics}
\label{table:meanNDCGSenti}
\end{table}

\begin{table}
\small
\begin{center}
\begin{tabular}{ |c|c|c|c| } 
 \hline
 \textbf{Dataset} & \textbf{p-value}  &  \textbf{\textit{Is Statistically Significant?}}\\ 
 \hline
 DOCLevel\_MeanAvgScore & 0.9515 &	\textbf{NO}\\ 
 \hline
 SENTLevel\_MeanAvgScore & 0.9441 &	\textbf{NO}\\  
 \hline
 ASPLevel\_MeanAvgScore & 0.3775 &	\textbf{NO}\\ 
 \hline
 \hline
 DOCLevel\_AvgNDCGScore & 0.6225 &	\textbf{NO}\\ 
 \hline
 SENTLevel\_AvgNDCGScore & 0.2562 &	\textbf{NO}\\ 
 \hline
 ASPLevel\_AvgNDCGScore & 0.1841 &	\textbf{NO}\\ 
 \hline
\end{tabular}
\end{center}
\caption{Two-tail t-test p-values Over All Controversial \\ Topics}
\label{table:pValues}
\end{table}

\subsubsection{NDCG-Senti Metric}
In addition to the \textit{sentiment polarity} metric, we also needed a more informative metric that will be convenient to compare particularly the retrieved document sets of two search engines. For this reason, we introduced a new metric, \textit{NDCG-Senti} that blends ranking and sentiment information, which is a variant of commonly used NDCG metric. We utilized the traditional formula of NDCG denoted in NDCG-FORMULA. For our ranking-sentiment metric \textit{NDCG-Senti}, we replaced relevance score \textit{$rel_i$} with sentiment score \textit{$pol_i$} in the formula.

With a slight modification in the traditional formula of NDCG, we came up with a modified NDCG scoring function which also takes document rank into consideration while comparing document sets of search engines.
However, there is still one more issue in the computation of transformed NDCG function that relevance grades in the traditional NDCG metric are always positive, whereas the real sentiment values as well as the transformed ones become negative since the polarity values obtained from TextBlob lie between -1 and 1. Note that the polarity values of documents were firstly transformed and then they were used for the computation of this second metric. In short, we computed \textit{NDCG-Senti} scores with transformed and subsequently normalized sentiment scores of TextBlob. The normalization was applied as follows: Transformed document polarity values of each query were normalized to the range of [0,1], then our new metric was obtained by using the non-negative sentiment polarities in each query. 

\begin{figure}
\centering
  \includegraphics[width=80mm]{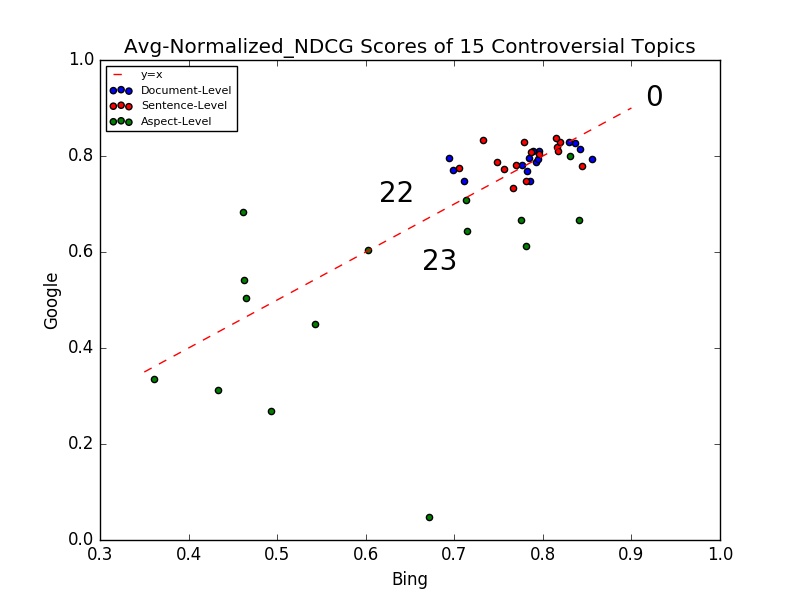}
  \caption{Avg-Normalized NDCG Scores of 15 Controversial Topics}
  \label{fig:avgNDCGScore}
\end{figure}

In order to get rid of the negative values, we utilize min-max normalization technique to map the polarities onto the range of [0, 1]. For each query, there are 10 transformed sentiment polarities of the documents returned from each search engine. We compute a minimum and a maximum value from these polarity values of the document set of each query. Then, we normalize the polarity values and obtain non-negative scores for the given set of documents. Subsequently, we compute a \textit{NDCG-Senti} score for the given set of documents with transformed sentiment polarities. In other words, we compute different min-max value pairs for each document set and normalize the polarity values  for each query separately. Then we computed an average \textit{NDCG-Senti} score for each controversial topic by using the \textit{NDCG-Senti} score of each query in the corresponding topic. You can find the table of mean average \textit{NDCG-Senti} scores in Table \ref{table:meanNDCGSenti} and the comparison plot of two search engines in terms of average \textit{NDCG-Senti} scores is displayed in Figure \ref{fig:avgNDCGScore}.

In addition to the comparative analysis in terms of average \textit{NDCG-Senti} scores for 15 controversial topics, we also compared two search engines by using \textit{NDCG-Senti} scores of each query in a controversial topic. In this analysis,
we compared two search engines per controversial topic and put the number of instances above/below and on the y=x axis to show the tendencies properly. While the previous analyses give the overall orientation of Bing \& Google on the whole dataset, per-controversial topic plots can elaborately depict the differences in their attitudes towards the given topics. You can find sample results per each controversial topic below.

%When one examines these per-controversial topic plots elaborately, it can be seen that in most of the controversial topics, neither of the two search engines shows a different attitude, except \textit{Medical Marijuana} and \textit{Syrian Refugees}. In \textit{Medical Marijuana}, Bing seems to be more positive, whereas in \textit{Syrian Refugees} Google tends to be more positive. Apart from these two controversial topics, the plots does not show explicit differences between these two search engines. 

\subsubsection{Statistical Significance Test}
After having computed and plotted sentiment and \textit{NDCG-Senti} scores in Figure \ref{fig:avgPolScore} and \ref{fig:avgNDCGScore} to understand the overall tendencies of two search engines, we also need to have a significance test to check if these differences are \textit{statistically significant}. For this, we applied the following procedure:

\begin{itemize}
\item For each given data sample, check that it is \textit{normally distributed} by creating a histogram from the data points.
\begin{itemize}
\item If so, then go to next step.
\item If not, transform the data points by using "log", "exp" etc.
\begin{itemize}
\item Check that if the given data sample is \textit{normally distributed} by creating a histogram from the transformed data points.
\item If not, try with a different transformation function for normality check and use other normality check methods such as QQplot until you observe that the data is \textit{normally distributed}.
\end{itemize}
\end{itemize}
\item Making sure that both of the data samples (Bing \& Google) are \textit{normally distributed}, apply F-test to see that the two data distributions have equal or unequal variances.
\item Apply two-tail t-test with equal or unequal variances to understand if the means of these two distributions are significantly different.
\end{itemize}

You can find the two-tail p values with .95 confidence level in Table \ref{table:pValues}.

\begin{figure}
\centering
  \includegraphics[width=80mm]{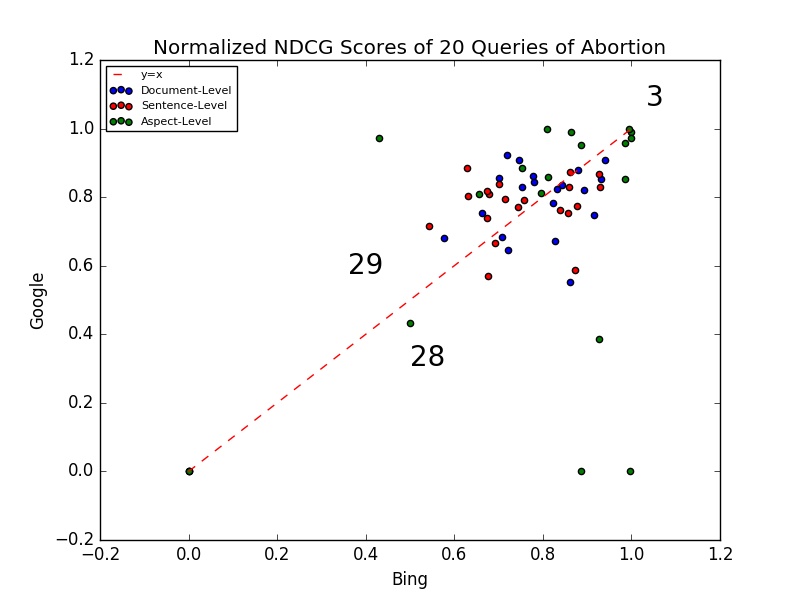}
  \caption{Normalized NDCG Scores of Abortion}
  \label{fig:NDCGScoreAbortion}
\end{figure}

\begin{figure}
\centering
  \includegraphics[width=80mm]{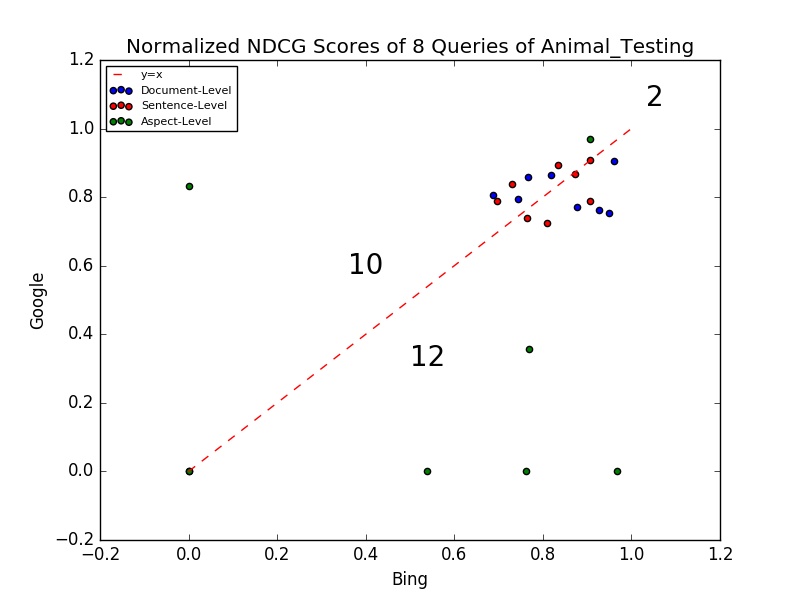}
  \caption{Normalized NDCG Scores of Animal Testing}
  \label{fig:NDCGScoreAnimalTesting}
\end{figure}

\begin{figure}
\centering
  \includegraphics[width=80mm]{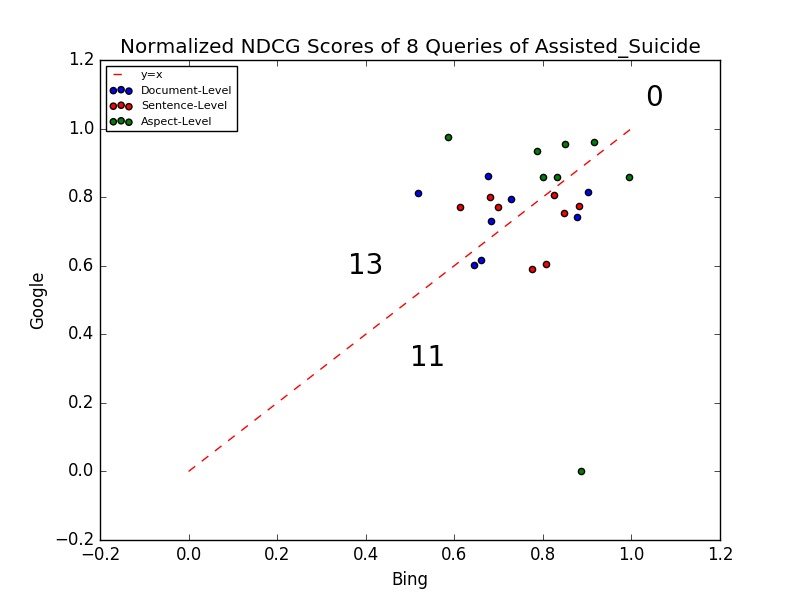}
  \caption{Normalized NDCG Scores of Assisted Suicide}
  \label{fig:NDCGScoreAssistedSuicide}
\end{figure}

\begin{figure}
\centering
  \includegraphics[width=80mm]{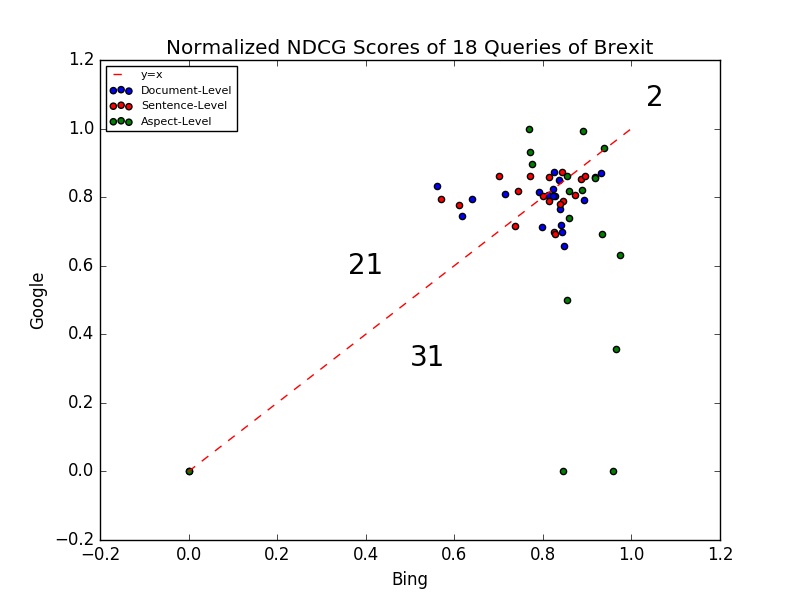}
  \caption{Normalized NDCG Scores of Brexit}
  \label{fig:NDCGBrexit}
\end{figure}

\begin{figure}
\centering
  \includegraphics[width=80mm]{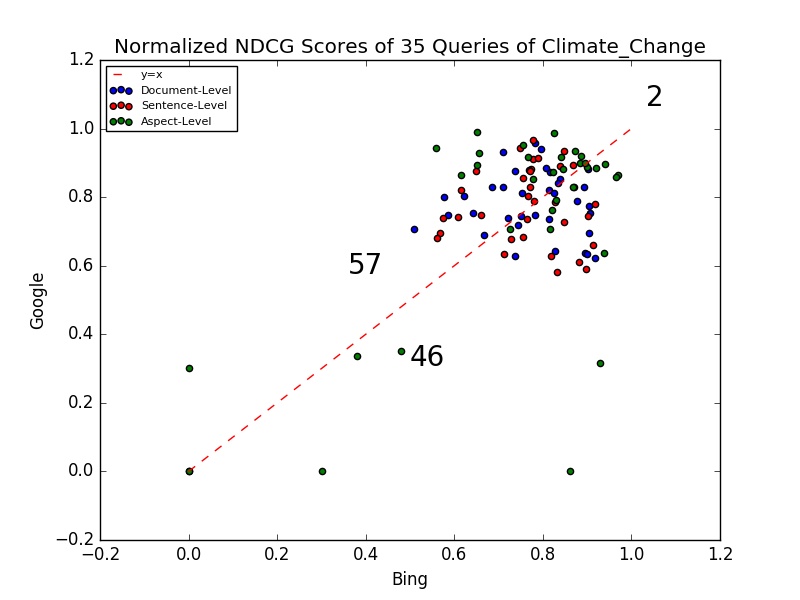}
  \caption{Normalized NDCG Scores of Climate Change}
  \label{fig:NDCGClimateChange}
\end{figure}

\begin{figure}
\centering
  \includegraphics[width=80mm]{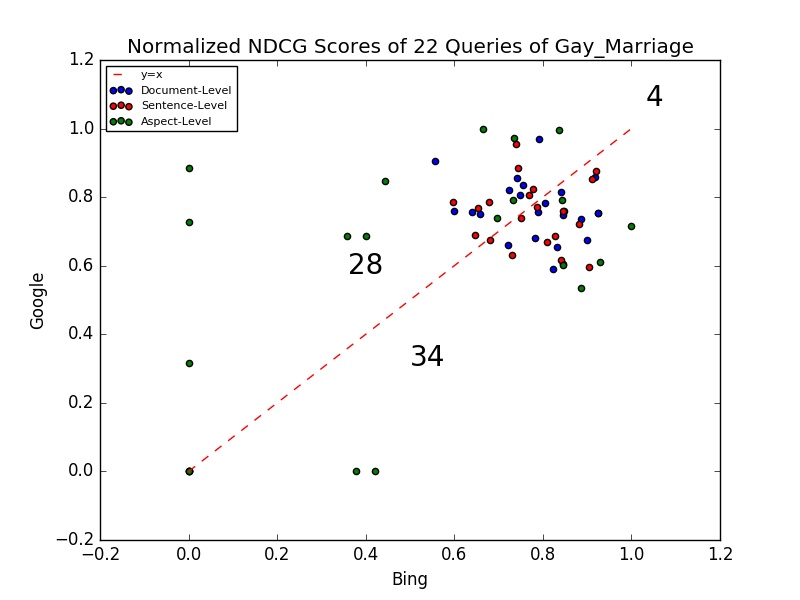}
  \caption{Normalized NDCG Scores of Gay Marriage}
  \label{fig:NDCGScoreGayMarriage}
\end{figure}

\begin{figure}
\centering
  \includegraphics[width=80mm]{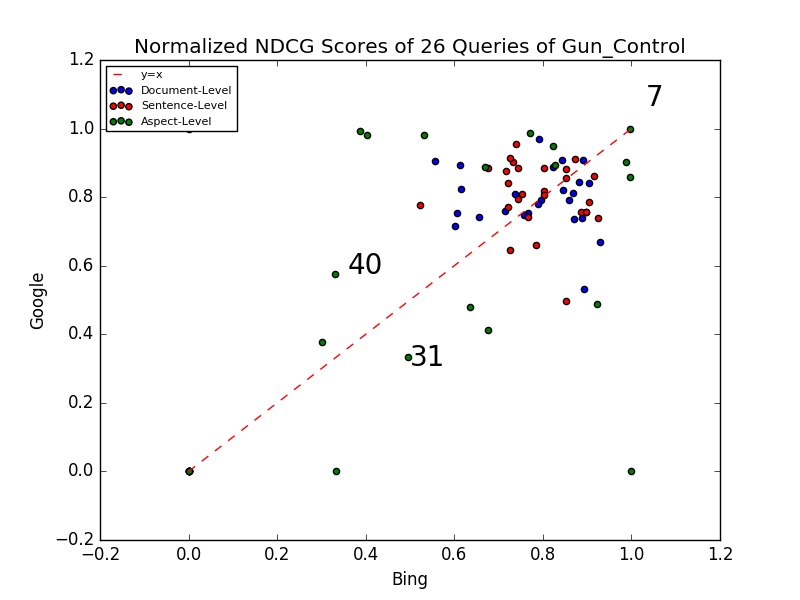}
  \caption{Normalized NDCG Scores of Gun Control}
  \label{fig:NDCGScoreGunControl}
\end{figure}

\begin{figure}
\centering
  \includegraphics[width=80mm]{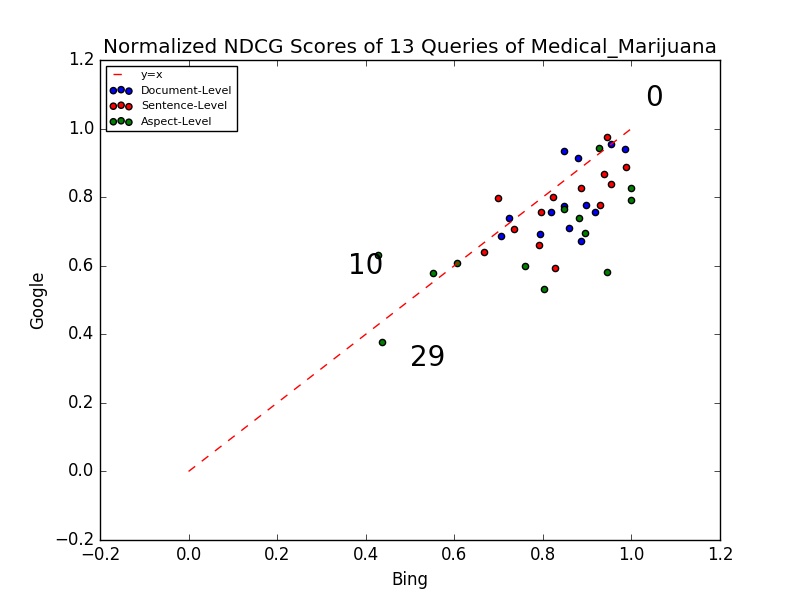}
  \caption{Normalized NDCG Scores of Medical Marijuana}
  \label{fig:NDCGMedicalMarijuana}
\end{figure}

\begin{figure}
\centering
  \includegraphics[width=80mm]{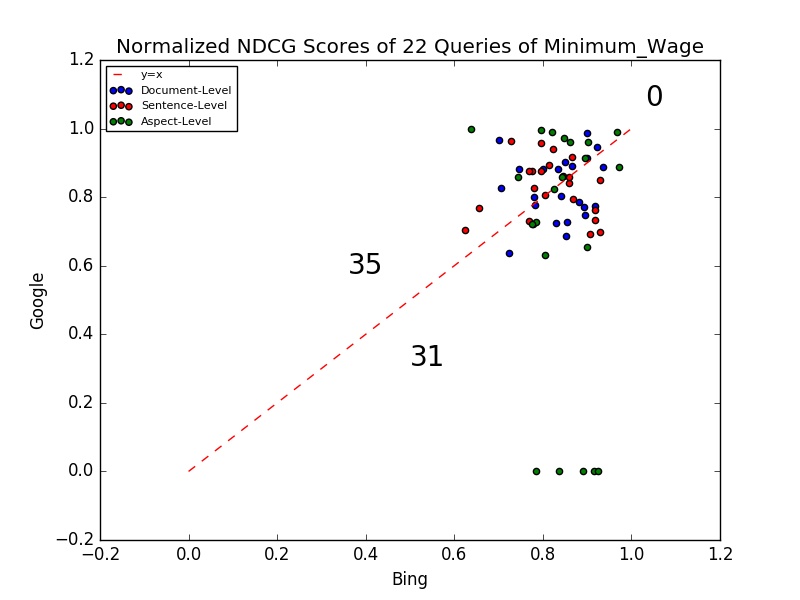}
  \caption{Normalized NDCG Scores of Minimum Wage}
  \label{fig:NDCGMinimumWage}
\end{figure}

\begin{figure}
\centering
  \includegraphics[width=80mm]{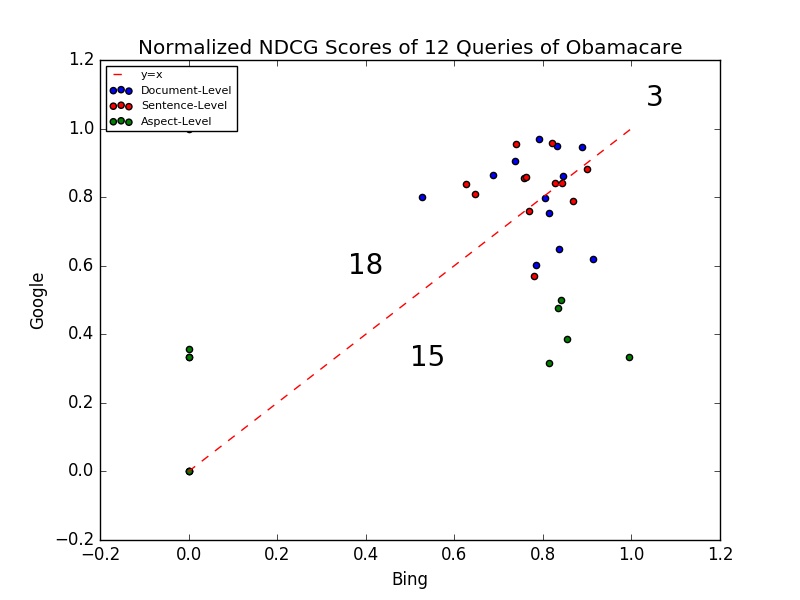}
  \caption{Normalized NDCG Scores of Obamacare}
  \label{fig:NDCGObamacare}
\end{figure}

\begin{figure}
\centering
  \includegraphics[width=80mm]{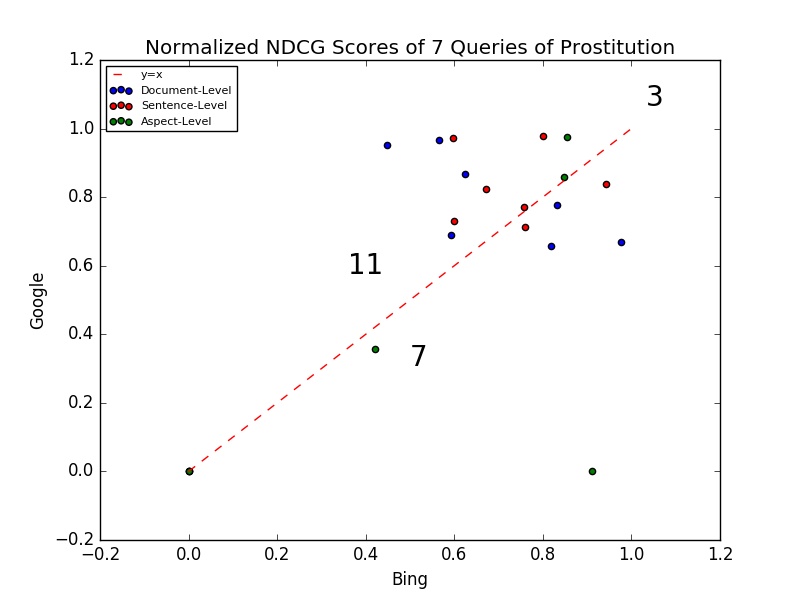}
  \caption{Normalized NDCG Scores of Prostitution}
  \label{fig:NDCGProstitution}
\end{figure}

\begin{figure}
\centering
  \includegraphics[width=80mm]{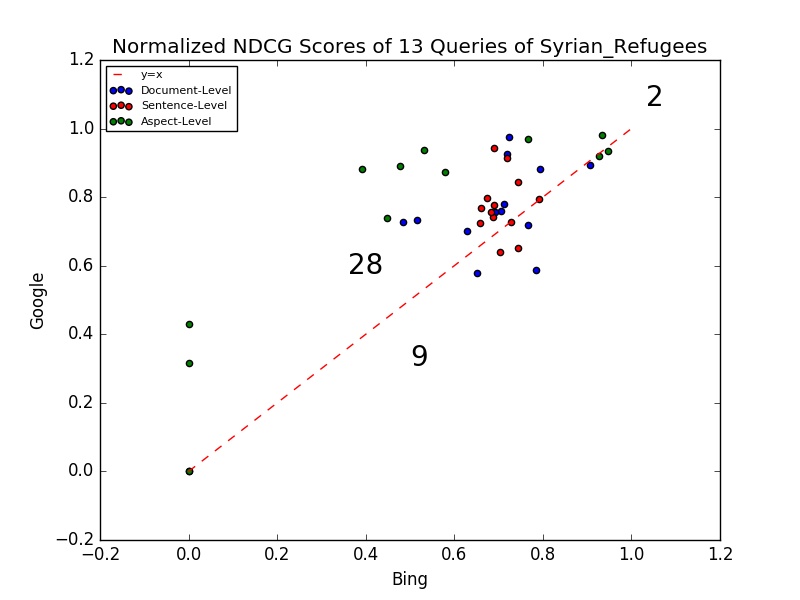}
  \caption{Normalized NDCG Scores of Syrian Refugees}
  \label{fig:NDCGScoreSyrianRefugees}
\end{figure}

\begin{figure}
\centering
  \includegraphics[width=80mm]{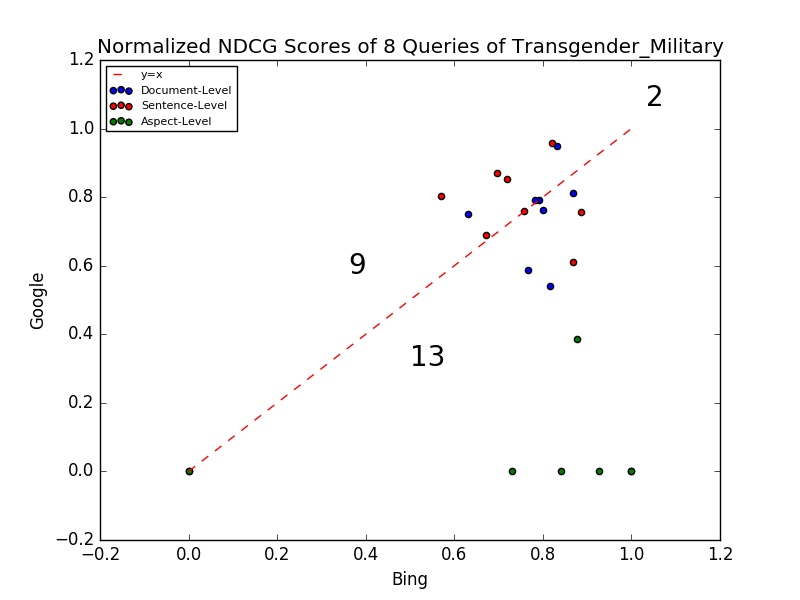}
  \caption{Normalized NDCG Scores of Transgender Military}
  \label{fig:NDCGScoreTransgenderMilitary}
\end{figure}

%I put those last two figures side by side with the code below!!!!
%%%%%%%%%%%%%%%%%%%%%%%%%%%%%%%%%%%%%%%%%%%%%

%%%%%%%%%%%%%%%%%%%%%%%%%%%%%%%%%%%%%%%%%%%%
%I put those last two figures side by side with the code below!!!!

\begin{figure}
  \centering
  \begin{minipage}[b]{0.45\textwidth}
    \includegraphics[width=\textwidth]{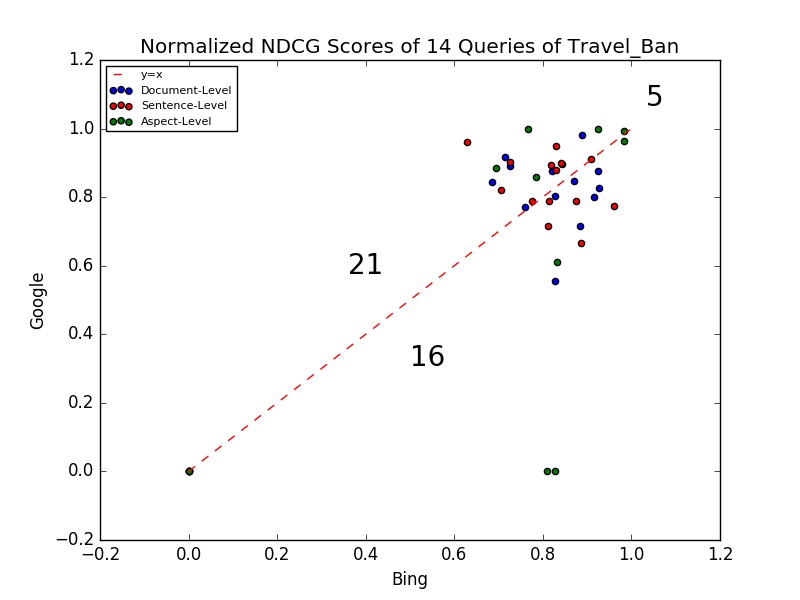}
    \caption{Normalized NDCG Scores of Travel Ban}
    \label{fig:NDCGScoreTravelBan}
  \end{minipage}
  \hfill
  \hfill
  \noindent
  \noindent
  \begin{minipage}[b]{0.45\textwidth}
    \includegraphics[width=\textwidth]{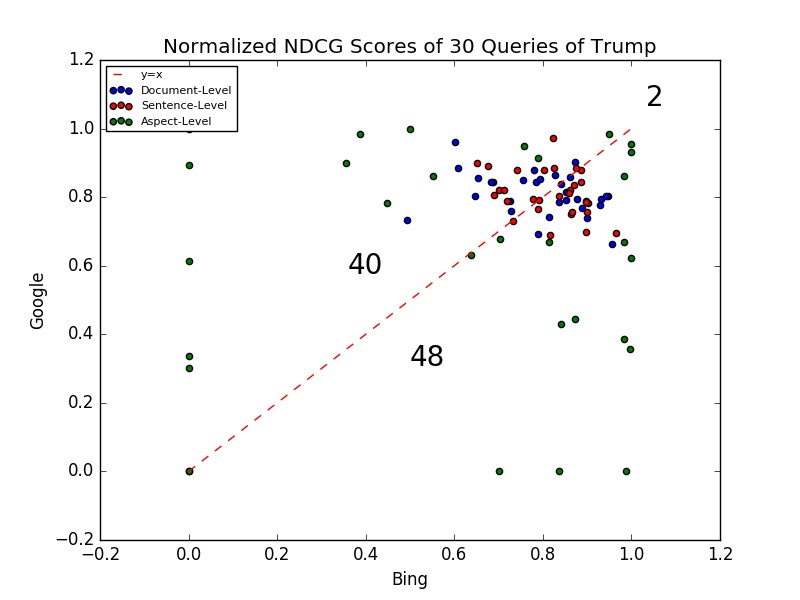}
    \caption{Normalized NDCG Scores of Trump}
    \label{fig:NDCGScoreTrump}
  \end{minipage}
\end{figure}

%At this point, we decided to compute \textit{NDCG-Senti} scores in two different ways by using original and only-positive set of sentiment scores and compare of these two results, as well. For each query, we directly computed the transformed NDCG function with original document polarities which also include negative values and the comparison plot at three distinct levels are shown in Figure [FIG-OriginalCompNDCG].

%Subsequently, we computed our transformed metric also with non-negative sentiment polarities for the document set of queries, and to get rid of negative values before computing the \textit{NDCG-Senti} score, min-max normalization technique was used to transform the polarities into the range of [0, 1]:
%\begin{equation}\label{eq:1}

%MinMax = \frac{($val_i$ - min)(max - min)}

%\end{equation}
%\prod_{i\in~Levels}{\frac{|workers_{maj}|}{|workers_i|}*\frac{|workers_
%\includegraphics{sentCompPlot}

\section{Future Work}
For the effective crowd labelling framework, we are waiting for the acceptance. If the paper is accepted, then we will extend the existing framework. For the second part, we are preparing the results for conference publication. Afterwards, that part can be hopefully extended for journal publication.

%\begin{acks}
 % The authors would like to thank Alan Turing \grantsponsor{GS501100001809}{National Natural
   % Science Foundation of
 %   China}{http://dx.doi.org/10.13039/501100001809} under Grant
%  No.:~\grantnum{GS501100001809}{61273304}
%  and~\grantnum[http://www.nnsf.cn/youngscientsts]{GS501100001809}{Young
  %  Scientsts' Support Program}.

%\end{acks}

\newpage

\bibliographystyle{ACM-Reference-Format}
\bibliography{sigproc} 

\end{document}